\documentclass[journal]{IEEEtran}

\ifCLASSINFOpdf

\else

\fi

\usepackage{times}
\usepackage[utf8]{inputenc}
\usepackage[compatibility=false]{caption}  
\usepackage{float}
\usepackage{lipsum}
\usepackage{amsmath,amsthm,amssymb,amsfonts}
\usepackage{makecell}
\usepackage{comment}
\usepackage{multirow}
\usepackage{rotating, graphicx}
\usepackage{xcolor}
\usepackage{tabularx}
\usepackage{orcidlink}
\usepackage{ragged2e}
\usepackage{textcomp}
\usepackage{listings}
\usepackage{algorithm}
\usepackage{algpseudocode}
\usepackage{setspace}
\usepackage[normalem]{ulem}
\useunder{\uline}{\ul}{}
\usepackage{hyperref}
\usepackage{xcolor}
\usepackage{booktabs} 
\usepackage[utf8]{inputenc}
\usepackage{cite}

\begin{document}

\title{Resisting Quantum Key Distribution Attacks Using Quantum Machine Learning}

\author{
    Ali Al-Kuwari\IEEEauthorrefmark{1}\textsuperscript{\orcidlink{0009-0007-2312-5921}} (alal55457@hbku.edu.qa),
    Noureldin Mohamed\IEEEauthorrefmark{1}\textsuperscript{\orcidlink{0009-0001-4150-8690}} (nomo89098@hbku.edu.qa),
    Saif Al-Kuwari\IEEEauthorrefmark{1}\IEEEauthorrefmark{2}\textsuperscript{\orcidlink{0000-0002-4402-7710}} (smalkuwari@hbku.edu.qa),
    Ahmed Farouk\IEEEauthorrefmark{1}\IEEEauthorrefmark{3}\textsuperscript{\orcidlink{0000-0001-8702-7342}} (ahsalem@hbku.edu.qa),
    and Bikash K. Behera\IEEEauthorrefmark{4}\textsuperscript{\orcidlink{0000-0003-2629-3377}} (bikas.riki@gmail.com)
    
    \IEEEauthorblockA{\IEEEauthorrefmark{1}Qatar Center for Quantum Computing, College of Science and Engineering, Hamad Bin Khalifa University, Doha, Qatar}\\
    \IEEEauthorblockA{\IEEEauthorrefmark{2}Corresponding author}\\
    \IEEEauthorblockA{\IEEEauthorrefmark{3}Department of Computer Science, Faculty of Computers and Artificial Intelligence, Hurghada University, Hurghada, Egypt}\\
    \IEEEauthorblockA{\IEEEauthorrefmark{4}Bikash's Quantum (OPC) Pvt. Ltd., Mohanpur, WB, 741246 India}
}

\maketitle

\begin{abstract}
The emergence of quantum computing poses significant risks to the security of modern communication networks as it breaks today's public-key cryptographic algorithms. Quantum Key Distribution (QKD) offers a promising solution by harnessing the principles of quantum mechanics to establish secure keys. However, practical QKD implementations remain vulnerable to hardware imperfections and advanced attacks such as Photon Number Splitting and Trojan-Horse attacks. In this work, we investigate the potential of quantum machine learning (QML) to detect popular QKD attacks. In particular, we propose a Hybrid Quantum Long Short-Term Memory (QLSTM) model to improve the detection of common QKD attacks. By combining quantum-enhanced learning with classical deep learning, the model captures complex temporal patterns in QKD data, improving detection accuracy. To evaluate the proposed model, we introduce a realistic QKD dataset that simulates typical QKD operations, along with seven attack scenarios: Intercept-and-Resend, Photon-Number Splitting (PNS), Trojan-Horse attacks on the Random Number Generator (RNG), Detector Blinding, Wavelength-dependent Trojan-Horse, and Combined attacks. The dataset includes quantum security metrics such as Quantum Bit Error Rate (QBER), measurement entropy, signal and decoy loss rates, and time-based metrics, ensuring an accurate representation of real-world conditions. Our results demonstrate the promising performance of the quantum machine learning approach compared to traditional classical machine learning models, highlighting the potential of hybrid techniques to enhance the security of future quantum communication networks. The proposed Hybrid QLSTM model achieved an accuracy of 94.7.0\% after 50 training epochs, outperforming classical deep learning models such as LSTM, CNN, ANN, Random Forest, and RNN. However, our evaluation is conducted on a semi-realistic, simulation-generated decoy-state BB84 dataset with simplified models of attacks and device imperfections, so the reported performance should be interpreted as a proof-of-concept rather than a final assessment on field-deployed QKD systems.

\end{abstract}


\vspace{1em}

{\small
\textbf{Data Availability Statement:} Dataset supporting the findings of this study is available from the corresponding author upon reasonable request.  

\textbf{Funding Statement:} None.  

\textbf{Conflict of Interest Disclosure:} The authors declare that they have no known conflict of interest that could have appeared to influence the work reported in this paper.  }
\\
\\

%
\IEEEpeerreviewmaketitle

\section{Introduction}
\label{sec:Introduction}
\IEEEPARstart{T}{he} rise of quantum computing poses a significant challenge to the security of modern communication networks, particularly for encryption techniques based on mathematical problems. As quantum computers evolve, the infamous Shor’s algorithm will be able to break current asymmetric encryption techniques \cite{gyongyosi2019survey}. In response to these emerging threats, researchers have been developing Post-Quantum Cryptography (PQC) to create encryption algorithms that resist quantum attacks. PQC algorithms rely on complex mathematical structures, such as lattice-based, code-based, and multivariate polynomial cryptosystems, making them resistant to current quantum techniques. However, despite its promising security features, PQC remains reliant on mathematical problems, so its security is not provably unconditional. Given the rapid advances in computational capabilities, even these new cryptographic approaches could face vulnerabilities in the future \cite{kumar2021state}.

An alternative and more radical direction is the Quantum Key Distribution (QKD), which uses principles of quantum mechanics to generate keys that are theoretically secure against eavesdropping \cite{gyongyosi2019survey, kumar2021state}. Unlike classical key exchange methods, QKD benefits from the no-cloning theorem, which prevents the cloning of quantum states. This property allows Alice (the sender) and Bob (the receiver) to detect any eavesdropping attempt, as such an intrusion inevitably disturbs the transmitted quantum states \cite{imran2024quantum}. This makes QKD attractive for sectors such as government, finance, and critical infrastructure, where extremely high levels of security are required \cite{gyongyosi2019survey}.

In recent years, QKD has progressed from theoretical studies to experimental prototypes and early-stage deployments. However, integrating QKD into existing communication infrastructures presents challenges related to scalability, network architecture, and resource optimization \cite{cao2022evolution}. Although global-scale QKD networks are being envisioned \cite{khenicomprehensive}, real-world implementations remain vulnerable to practical threats. Hardware imperfections, channel noise, and side-channel attacks—such as Photon Number Splitting (PNS), Trojan Horse Attacks, and Channel Tampering—can undermine the theoretical guarantees of QKD \cite{khenicomprehensive, imran2024quantum}. These threats underscore the importance of robust monitoring and intrusion-detection strategies.

Researchers have explored both conventional and data-driven approaches to securing QKD against such vulnerabilities. On the traditional side, non-ML methods focus on signal analysis, hardware countermeasures, and statistical detection. For example, spectral estimation techniques have been used to detect low-rate denial-of-service (LDoS) attacks in CV-QKD by identifying distinct low-frequency patterns in the power spectral density (PSD) \cite{dai2022low}. Phase-sensitive amplifiers (PSA) combined with homodyne detection have been deployed to improve resilience against individual attacks \cite{alshaer2024enhancing}, while plug-and-play architectures for Differential Phase Shift Measurement-Device-Independent QKD (DPS-MDI-QKD) help mitigate vulnerabilities in state preparation and measurement \cite{sharma2024mitigating}. Network-layer defense strategies based on real-time traffic monitoring have also been proposed to detect denial-of-service attempts \cite{li2024detection}. Although these methods improve QKD robustness, they are often constrained by predefined assumptions, limiting their adaptability to novel or evolving threats.

This limitation has motivated a growing interest in Machine Learning (ML) and Deep Learning (DL) for QKD security. Unlike static threshold-based methods, ML models can adapt to dynamic patterns and detect known and previously unseen attacks. In Continuous-Variable QKD (CV-QKD), Du and Huang \cite{du2022multi} proposed multi-class detection using Binary Relevance Neural Networks (BR-NN) and Label Powerset Neural Networks (LP-NN), combined with one-class SVM for unknown attacks. Kish \emph{et al.} \cite{kish2024mitigation} developed a decision tree–based framework to detect and mitigate channel amplification attacks and related DoS scenarios. Mao \emph{et al.} \cite{mao2020detecting} used Artificial Neural Networks (ANN) to classify multiple attack types from optical parameter features such as quadrature values and local oscillator intensity. Luo \emph{et al.} \cite{luo2022beyond} introduced a semi-supervised GAN-based method that improved the detection generalization of unseen threats. Table~\ref{tab:cvqkd_ml_attacks} summarizes these CV-QKD approaches.

\begin{table*}[!t]
    \centering
    \caption{Detecting quantum attacks in CV-QKD via ML}
    \label{tab:cvqkd_ml_attacks}
    \begin{tabular}{|p{2cm}|p{2.5cm}|p{2.5cm}|p{2.5cm}|p{2.5cm}|} 
        \hline
        & \cite{luo2022beyond} & \cite{mao2020detecting} & \cite{du2022multi} & \cite{kish2024mitigation} \\ \hline
        \textbf{Training Data} & Experimental data from a real CV-QKD system. & Simulated CV-QKD parameters. & Simulated CV-QKD parameters. & Estimated CV-QKD parameters. \\ \hline
        \textbf{Attacks Covered} & Calibration, LO intensity, saturation, wavelength, unknown threats. & Calibration, LO intensity, saturation, hybrid attacks. & Calibration, LO intensity, saturation. & Channel Amplification (CA), CA-DoS, DoS. \\ \hline
        \textbf{Model} & GAN with anomaly detection. & ANN. & BR-NN and LP-NN + one-class SVM. & Decision tree. \\ \hline
        \textbf{Accuracy} & 99.3\%. & $>$99\%. & 100\% (known), 98.7--99.8\% (unknown). & 100\% (low noise), 90.1\% (high noise). \\ \hline
        \textbf{Generaliza-tion} & High & Moderate & High & Low \\ \hline
        \textbf{Advantages} & Strong generalization. & Effective multi-attack classification. & Known + unknown detection. & Fast classification. \\ \hline
        \textbf{Limitations} & High complexity. & Reduced SKR and distance. & Potential false positives. & Narrow training scope. \\ \hline
    \end{tabular}
\end{table*}

In Discrete-Variable QKD (DV-QKD), similar ML-based methods have been explored. Al \emph{et al.} \cite{al2021machine} applied ANN and LSTM models to detect Man-in-the-Middle attacks in QKD networks integrated with IoT. Tunc \emph{et al.} \cite{tunc2023machine} used LSTM and SVM classifiers to detect eavesdropping in quantum communication channels. Xu \emph{et al.} \cite{xu2024automatically} implemented Random Forest classifiers capable of detecting both device imperfections and eavesdropping in real-time. Table~\ref{tab:dvqkd_ml_attacks} summarizes these DV-QKD approaches.

\begin{table}[!t]
    \centering
    \caption{Detecting quantum attacks in DV-QKD via ML}
    \label{tab:dvqkd_ml_attacks}
    \resizebox{\columnwidth}{!}{  
    \begin{tabular}{|p{3cm}|p{3cm}|p{3cm}|p{3cm}|}
        \hline
        & \cite{al2021machine} & \cite{tunc2023machine} & \cite{xu2024automatically} \\ \hline
        \textbf{Training Data} & Simulated QKD key length data. & Theoretical BB84 attack simulations. & Experimental + theoretical data. \\ \hline
        \textbf{Attacks Covered} & Man-in-the-middle. & Eavesdropping. & Device imperfections + eavesdropping. \\ \hline
        \textbf{Model} & ANN, LSTM. & SVM, LSTM. & RF. \\ \hline
        \textbf{Accuracy} & 99.1\%. & 100\%. & 98\%. \\ \hline
        \textbf{Generalization} & Low & Low & High \\ \hline
        \textbf{Advantages} & Effective for predefined attacks. & Accurate detection. & Detects known + unknown threats in real-time. \\ \hline
        \textbf{Limitations} & Simple assumptions, high computation time. & Limited scope. & High complexity. \\ \hline
    \end{tabular}
    }
\end{table}

Despite these advances, existing ML approaches often rely on idealized simulations and limited attack scenarios. They tend to omit realistic conditions such as channel loss, phase noise, photon shot noise, depolarization, and detector imperfections, and often ignore critical quantum-level threats such as PNS, Intercept-and-Resend, Trojan Horse, RNG, detector blinding, and Wavelength-Dependent attacks. Furthermore, to the best of our knowledge, Quantum Machine Learning has not yet been applied to QKD intrusion detection.

To bridge this gap, we propose a hybrid intrusion detection system that integrates Quantum Long Short-Term Memory (QLSTM) into QKD monitoring. By combining quantum-enhanced learning with classical deep learning, the proposed model aims to improve detection accuracy and adaptability to evolving threats. We also introduce a new simulated dataset for the decoy-state BB84 DV-QKD under eight scenarios: normal operation, Intercept-and-Resend, Trojan Horse, PNS, RNG, Wavelength-dependent Trojan Horse, Detector Blinding, and combined attacks.

\subsection{Contributions} 
The contributions of this work can be summarized as follows:

\begin{itemize}
    \item \emph{Comprehensive QKD dataset:} Given the lack of publicly available and comprehensive datasets, this work proposes a realistic QKD simulation using the PennyLane library. The dataset simulates decoy-state BB84 DVQKD under eight different scenarios: normal QKD operation, Intercept-and-Resend attack, Trojan Horse attack, PNS attack, RNG attack, Wavelength-dependent Trojan Horse attack, Detector Blinding attack, and combined attack. It also includes various quantum metrics such as Quantum Bit Error Rate (QBER), measurement entropy, signal and decoy loss rates, and time-based metrics, ensuring a dataset that accurately reflects real-world conditions.
    \item \emph{Quantum Machine Learning for quantum intrusion detection on QKD systems:} A hybrid intrusion detection model is proposed, integrating a QLSTM network with a classical LSTM. This approach leverages the strengths of both quantum computation and classical deep learning, offering a practical intrusion detection solution for future QKD networks. 
\end{itemize}

\subsection{Organization}
The remainder of this paper is organized as follows. Section~\ref{sec:Model} introduces the proposed Hybrid QLSTM model to detect QKD attacks. Section~\ref{sec:Dataset} discusses the dataset generation process, including the simulation algorithm that was used to generate the dataset. Section~\ref{sec:Evaluation} illustrates the evaluation setup and the metrics used. Section~\ref{sec:Results} presents and analyzes the experimental results, focusing on the performance of the proposed hybrid QLSTM model. Finally, Section~\ref{sec:Section_7} concludes this paper and provides suggestions for future work.

\section{Model}
\label{sec:Model}

This section presents the Hybrid QLSTM Model for QKD attack detection. Given the sequential nature of QKD transmissions, QLSTM is used to analyze quantum security metrics over time, thereby improving attack detection accuracy compared to classical LSTM. The model consists of a QLSTM layer, a classical LSTM layer, and a fully connected layer. The QLSTM captures quantum-enhanced sequential features, the LSTM refines temporal dependencies, and the final layer maps features to classification logits. The section also covers GPU-accelerated training using CrossEntropyLoss, AdamW optimization, and a cosine annealing learning rate schedule. The model is trained for 50 epochs, with early stopping to prevent overfitting.

In this work, we use QLSTM to analyze QKD security metrics over time and detect anomalies. QLSTM is an extension of the classical Long Short-Term Memory (LSTM) model that incorporates quantum principles to enhance learning efficiency in sequential data tasks. Traditional LSTMs are widely used in time-series analysis, making them well-suited to detect patterns in QKD-based attack scenarios. However, QLSTM leverages quantum mechanics to capture dependencies more effectively than classical models. As shown in \cite{tripathi2025quantum}, QLSTMs outperformed classical LSTMs in detecting anomalies in Distributed Denial of Service (DDoS) attacks. Given the sequential nature of QKD transmissions, detecting anomalies in quantum metrics is crucial. Compared to classical LSTMs, QLSTMs converge faster and achieve higher accuracy in detecting attack patterns, making them a strong candidate for quantum security applications.

\subsection{Foundations of LSTM and QLSTM}
Traditional ML models, such as LSTM, are designed to process sequential data by learning temporal dependencies through gated recurrent units. LSTM networks have been successfully applied in various anomaly-detection and time-series tasks; however, their representational capacity is bounded by classical computational limits. QML extends these methods by embedding data in quantum states and processing them through variational quantum circuits (VQCs). These circuits exploit quantum mechanical principles, such as superposition, entanglement, and interference, to represent and manipulate complex correlations in fewer dimensions. This provides a potential advantage in optimization efficiency. The QLSTM model combines these properties and replaces or augments the classical LSTM gates with variational quantum circuits, enabling the model to learn temporal relationships within a quantum-enhanced feature space. This hybrid structure allows quantum layers to capture complex dependencies in the input sequence. In contrast, classical layers handle large-scale temporal aggregation, thereby improving convergence and generalization in sequential learning tasks.

Quantum circuits operate within a high-dimensional Hilbert space, where qubit superposition and entanglement allow data to be represented in exponentially richer feature spaces compared to classical vector encoding \cite{khan2024quantum}. This enables more expressive temporal feature extraction with fewer parameters, leading to improved optimization landscapes and enabling faster gradient-based training. Furthermore, many studies have shown that QLSTM models exhibit accelerated convergence and enhanced prediction accuracy in various time-series domains, including renewable energy forecasting \cite{khan2024quantum} and cybersecurity \cite{tripathi2025quantum}. In these studies, QLSTM consistently outperformed its classical LSTM counterpart, achieving lower validation loss within early epochs while maintaining superior generalization on unseen samples. These results empirically confirm that quantum recurrent layers can mitigate vanishing-gradient effects and capture complex nonlinear dependencies more effectively than purely classical architectures.

\subsection{Model Architecture}
\begin{figure*}[!t]
    \centering
    \includegraphics[width=1.8\columnwidth]{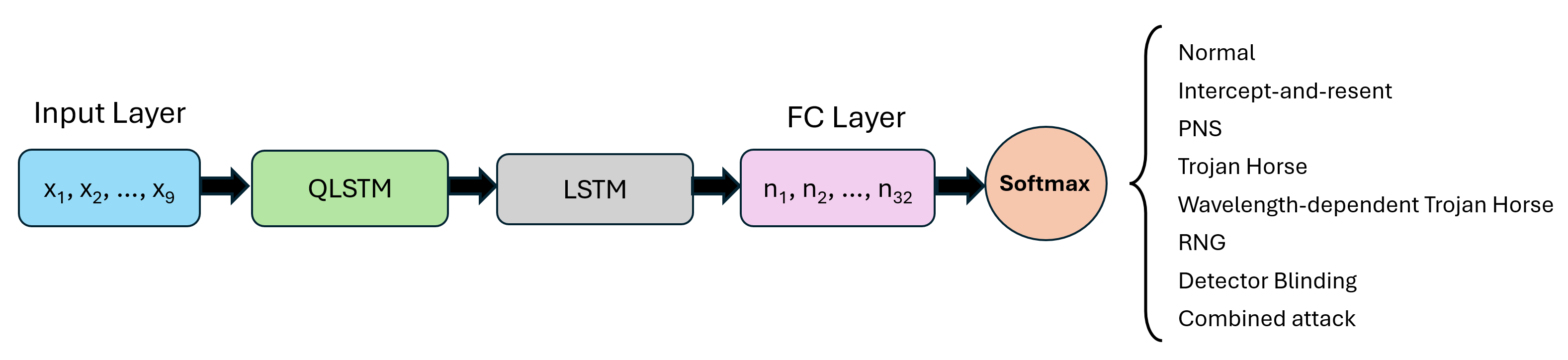}
    \captionsetup{justification=centering}
    \caption{High Level Diagram of The Proposed Hybrid QLSTM Model}
    \label{fig:qlstm_diagram}
\end{figure*}

The model consists of three main layers: a QLSTM layer, a classical LSTM layer, and a fully connected layer. Figure~\ref{fig:qlstm_diagram} shows the high-level design of the proposed model. The QLSTM layer processes the sequential input data while incorporating quantum properties, such as entanglement and superposition, capturing complex features in the data. The output of the QLSTM layer is fed into a classical LSTM layer, which learns the temporal patterns from the quantum-enhanced sequences. Finally, a fully connected layer maps the learned features to the output space, producing logits, which are the scores that serve as the output of an ML model. These logits are passed to a softmax function for classification. Pseudocode~\ref{alg:qlstm} describes the model.

\begin{algorithm}[]
    \caption{Hybrid QLSTM Model}
    \label{alg:qlstm}
    \begin{algorithmic}[1]
    
        \State \textbf{1. Quantum LSTM (QLSTM) Layer Initialization}
            \State \indent Set parameters: 
                \begin{itemize}
                    \item $input\_size$
                    \item $hidden\_size$
                    \item $n\_qubits = 9$
                    \item $n\_qlayers = 1$
                    \item $backend = $ ``default.qubit''
                \end{itemize}
            \State \indent Use Strongly Entangling Layers for quantum gates
            \State \indent Initialize quantum circuit weights and entanglement parameters
            \State \indent Define quantum gates for input, forget, output, and candidate cell states
            \State \indent Post-process quantum measurements for classical compatibility

        \State \textbf{2. Classical LSTM Layer Initialization}
            \State \indent Set input size $= n\_qubits = 9$
            \State \indent Set hidden size $= 32$
            \State \indent Initialize weights and hidden/cell states

        \State \textbf{3. Fully Connected (FC) Layer}
            \State \indent Define linear layer mapping $32 \rightarrow output\_dim = 4$

        \State \textbf{4. Forward Pass}
        \Function{Forward}{$x$}
            \State \textbf{Input:} $x$ of shape $(batch\_size, seq\_len, feature\_dim)$
            \State Initialize $h_0, c_0$ for QLSTM as zeros
            \For{each time step $t$ in $seq\_len$}
                \State $x_t \gets x[:, t, :]$ \Comment{Extract features at time $t$}
                \State $h_t, c_t \gets \text{QLSTM}(x_t, h_{t-1}, c_{t-1})$ \Comment{Quantum processing}
                \State Store $h_t$
            \EndFor
            \State $qlstm\_output \gets$ Stack all $h_t$ along time axis
            \State $lstm\_output, (h_n, c_n) \gets \text{ClassicalLSTM}(qlstm\_output)$
            \State $last\_hidden \gets h_n$ \Comment{Or $lstm\_output[:,-1,:]$}
            \State $logits \gets \text{FC}(last\_hidden)$
            \State \Return logits
        \EndFunction

    \end{algorithmic}
\end{algorithm}

\begin{figure*}[!t]
    \centering
    \includegraphics[width=\textwidth]{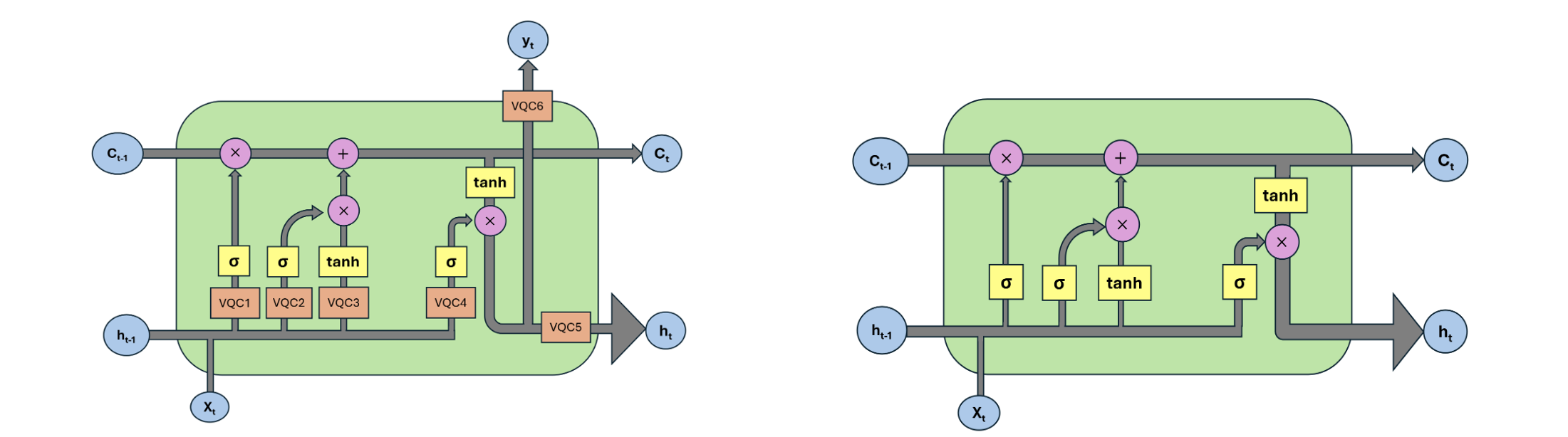}
    \caption{(a) QLSTM and (b) LSTM Model Architectures}
    \label{fig:combined_qlstm_lstm}
\end{figure*}

\subsubsection{QLSTM Layer}
 This layer is implemented based on the open-source QLSTM implementation in \cite{qlstm_github}. However, for the purpose of this project, the original entanglement layer was changed from Basic Entanglement to Strongly Entanglement Layer. At each time step, the QLSTM receives the current input $x_t$ and the previous hidden and cell states, $h_t$ and $c_t$, respectively. It processes this information using VQCs that represent the Forget Gate, the Input Gate, the Candidate Cell State, and the Output Gate. Each VQC operates on qubits that use superposition and entanglement to encode information in a high-dimensional Hilbert space. This enables the gates to represent and learn complex feature relationships with fewer parameters than classical layers, improving training efficiency and representation power. Specifically, the gates are computed as follows:

{\small
\begin{align}
    i_t &= \sigma( VQC_2(v_t) ) \quad 
    f_t = \sigma( VQC_1(v_t) ), \\ 
    \tilde{C}_t &= \tanh( VQC_3(v_t) ), \quad
    o_t = \sigma( VQC_4(v_t) ), 
\end{align}
}

where, $v_t$ is the $N$-dimensional input vector at time step $t$, $VQC_k(\cdot)$ represents a parameterized variational quantum circuit for gate $k$, $\sigma$ denotes the sigmoid activation function that regulates information flow, $\tanh$ denotes the hyperbolic tangent activation function for candidate states, and $i_t, f_t, \tilde{C}_t, o_t$ are the input, forget, candidate cell state, and output gates at time $t$, respectively.

The outputs of the VQCs are passed through classical nonlinearities ($\sigma$, $\tanh$) before combining to produce the updated hidden and cell states. These updated states are then propagated to the next time step. The architecture of the QLSTM model used in this work is visually illustrated in Figure~\ref{fig:combined_qlstm_lstm} (a).

\subsubsection{Classical LSTM Layer}
The LSTM captures long-term and short-term dependencies in sequential data through its cell and hidden states. 
It regulates information flow using input, forget, and output gates, which decide what to keep or discard from previous time steps. 
In this model, the LSTM refines the temporal features extracted from the QLSTM, ensuring stable learning and smoother transitions before classification.
 In this proposed model, the LSTM layer consists of 32 hidden states and operates in three main steps:

\paragraph{Input Processing} The quantum layer's output is reshaped into a sequence format to match the LSTM's expected input $x_t$.

\paragraph{LSTM Processing} At each time step $t$, the LSTM layer updates its internal memory using input $x_t$ and the previous hidden and cell states ($h_{t-1}$, $c_{t-1}$). As illustrated in Figure~\ref{fig:combined_qlstm_lstm} (b), this process involves three gates, the forget gate $f_t$, the input gate $i_t$, and the output gate $o_t$, along with a candidate cell state $\tilde{c}_t$. Gate activations are computed as follows:
    \begin{align}
        i_t &= \sigma \big( W_{xi} x_t + W_{hi} h_{t-1} + b_i \big), \\
        f_t &= \sigma \big( W_{xf} x_t + W_{hf} h_{t-1} + b_f \big), \\
        \tilde{c}_t &= \tanh \big( W_{xc} x_t + W_{hc} h_{t-1} + b_c \big), \\
        o_t &= \sigma \big( W_{xo} x_t + W_{ho} h_{t-1} + b_o \big).
    \end{align}
where, $\sigma(\cdot)$ is the sigmoid activation function, $\tanh(\cdot)$ is the hyperbolic tangent activation function, $x_t$ is the current input vector, $h_{t-1}$ and $c_{t-1}$ are the hidden and cell states from the previous time step, $i_t$, $f_t$, and $o_t$ are the input, forget, and output gate activations, respectively, $\tilde{c}_t$ is the candidate cell state, $W_{x\ast}$ and $W_{h\ast}$ are the learnable weight matrices, and $b_\ast$ are the bias vectors for each gate.

\paragraph{Hidden State Extraction} Once the sequence is fully processed, the last hidden state $h_t$ of the LSTM model is extracted to be used in the final prediction step.
    \begin{align}
        c_t = f_t \odot c_{t-1} + i_t \odot \tilde{c}_t,
    \end{align}
where $\odot$ denotes element-wise multiplication. Finally, the hidden state $h_t$ is updated as:
    \begin{align}
        h_t = o_t \odot \tanh(c_t).
    \end{align}    
This ensures that the output at each step is a filtered representation of the current cell state. After processing the entire sequence, the final hidden state $h_t$ is extracted and passed to the next stage for classification. 
This process allows the model to capture long-term temporal dependencies while mitigating vanishing gradients.

\subsubsection{Fully Connected Layer}
The fully connected layer maps the classical LSTM's output to the final output space, which consists of 4 logits, each corresponding to a possible class (Normal, Intercept-and-Resend Attack, PNS Attack, Trojan-Horse Attack). These logits are later transformed to class probabilities using a softmax function during evaluation.

\section{Dataset}
\label{sec:Dataset}

This Section introduces a semi-realistic QKD dataset designed to address the limitations of existing datasets, which often rely on mathematical assumptions or lack specific attack scenarios. By incorporating PNS, intercept-and-resend, Trojan-horse, RNG, detector blinding, wavelength-dependent, and combined attacks, the dataset better represents real-world threats to quantum communication, though with some limitations. Key quantum metrics are collected to evaluate the security of the QKD process. The dataset goes through preprocessing steps, including feature selection, normalization, and augmentation, to ensure its suitability for the proposed hybrid QLSTM model. This dataset serves as a benchmark for evaluating the effectiveness of the QLSTM model in detecting QKD security threats.

\subsection{Threat Model}
\label{sec:threat_model}
We generate the dataset under a decoy-state BB84 DV-QKD adversarial framework. Eve controls the quantum channel, but has no physical access to Alice's/Bob's internals. She can (i) perform basis measurements and resend states (Intercept--Resend), (ii) select photon-number-dependent strategies (PNS), (iii) inject bright or off-wavelength light (Trojan-Horse variants), and (iv) drive detectors into a linear regime (Detector Blinding). Eve does not replace the quantum channel or perform loss compensation, nor does she calibrate the return flux (as in the standard THA), consistent with a bounded but capable adversary. These are side-channel and channel-level capabilities that appear through observable statistics (QBER, detection/loss rates, entropy, and timing). Consistent with \cite{hwang2003quantum, lydersen2010hacking, jain2016attacks, gisin2006trojan}, we implement simplified simulation-feasible variants that preserve the operational effect of each attack while avoiding full hardware modeling. Any deviations from the strongest known forms (e.g., loss-compensation in PNS or power-calibrated THA) are documented in each scenario as modeling limitations.

\subsection{Simulation Setup}
Existing QKD datasets (as discussed in Section~\ref{sec:Introduction}) often suffer from significant limitations, such as containing very few features, relying on mathematical assumptions, or focusing on the practical vulnerabilities of QKD systems. In addition, many studies focus on general eavesdropping threats without specifying the exact attack strategy or emphasizing device imperfections. In contrast, this work explicitly incorporates PNS, Intercept-and-Resend, and Trojan-Horse attacks, which directly exploit QKD vulnerabilities at the photon transmission level. The selection of these attacks is driven by the gaps identified in previous research. By addressing these threats, our dataset enables a more precise evaluation of ML-based detection mechanisms. 

In this work, we adopt the decoy-state BB84 protocol, which enhances security by randomly transmitting additional photon (decoy) pulses with varying intensities. Alice randomly sends both signal and decoy pulses, and Bob measures them as in the standard BB84 protocol. Since Eve cannot distinguish between signal and decoy pulses, she cannot selectively measure only multi-photon pulses without being detected. By analyzing the detection rates for both signal and decoy states, Alice and Bob can identify and mitigate PNS attacks before finalizing the key \cite{hwang2003quantum}.

The simulation of the decoy-state BB84 protocol includes eight scenarios: Normal, Intercept-and-Resend, PNS, Trojan-Horse, Wavelength-Dependent Trojan-Horse, RNG, Detector Blinding, and Combined attacks, consistent with the threat model in Section~\ref{sec:threat_model}. It collects various quantum metrics, including the Quantum Bit Error Rate (QBER), bit mismatch entropy, signal and decoy detection rates, and time-based statistics, as discussed in table~\ref{tab:qkd_metrics}. The simulation is carried out to generate approximately a 256-bit secret key while reflecting practical transmission conditions. To make the dataset more realistic, different noise sources are incorporated. In particular, quantum channel losses are introduced to simulate photon attenuation along the link, photon shot noise is added to reflect the inherent fluctuations in photon detection, and depolarizing noise is applied to model random state decoherence that naturally occurs during transmission. These noise sources vary randomly for each photon transmission, with their strength set to low, moderate, or high levels. Although this does not capture every possible imperfection in a real system, it provides a closer approximation to realistic conditions compared to purely ideal simulations. Table~\ref{tab:qkd_dist} summarizes the distribution of attack scenarios in the dataset, which contains 10,329 samples uniformly distributed on eight labels. The following subsections describe each scenario and explain how the impact of each condition was incorporated during dataset construction. Table \ref{tab:attack_conditions} summarizes the conditions we considered in each scenario.
\begin{table}[H]
    \centering
    \caption{Distribution of QKD Attack Scenarios}
    \label{tab:qkd_dist}
    \resizebox{\columnwidth}{!}{  

    \begin{tabular}{|p{5.5 cm}|c|}
        \hline
        \textbf{Scenario} & \textbf{ Samples} \\
        \hline
        Normal QKD  & 1292 \\
        Man-in-the-Middle (MITM) & 1291 \\
        Photon Number Splitting (PNS) & 1291 \\
        Trojan-Horse Attack & 1291 \\
        Wavelength Dependent Trojan-Horse & 1291 \\
        Random Number Generator Attack & 1291 \\
        Detector Blinding Attack & 1291 \\
        Combined Attack & 1291 \\
        \hline
    \end{tabular}
    }
\end{table}

\begin{table*}[!t]
    \centering
    \caption{Summary of Attack Scenarios and Conditions Considered}
    \label{tab:attack_conditions}
    \begin{tabular}{|p{3cm}|p{3cm}|p{3cm}|p{3cm}|}
        \hline
        \textbf{Scenario} & \textbf{Description} & \textbf{Key Conditions Considered} & \textbf{Expected Impact on QKD} \\
        \hline
        \textbf{Normal QKD} & Standard QKD process without any attack. Alice and Bob exchange a key securely. & No external interference, stable noise levels, and proper key sifting. & Low QBER, stable detection rates, and normal transmission times. \\
        \hline
        \textbf{Intercept-Resend Attack} & Eve intercepts and measures photons before resending them to Bob. & Eve randomly measures on a basis, causing transmission errors. & Increased QBER, mismatch entropy, and detection errors. \\
        \hline
        \textbf{Trojan-Horse Attack} & Eve injects additional photons into Alice’s system to extract internal settings. & Eve controls the intensity and timing of injected photons. & Potential bias in key generation and increased information leakage. \\
        \hline
        \textbf{Photon-Number Splitting (PNS) Attack} & Eve selectively measures multi-photon pulses and retains a copy of the key. & Eve detects photon pulses without altering single-photon transmissions. &Lower detection rates without a significant increase in QBER. \\
        \hline
        \textbf{Wavelength-Dependent Trojan-Horse Attack} & Eve injects pulses at off-wavelengths to exploit detector sensitivity. & Wavelength differences affect detection efficiency and phase encoding. & Increased QBER, timing delays, and reduced detection efficiency. \\
        \hline
        \textbf{Random Number Generator (RNG) Attack} & Eve exploits bias or predictability in Alice’s and Bob’s random number generators. & Biased or predictable bit and basis choices due to compromised RNGs. & Increased predictability of key bits, subtle bias in entropy, possible QBER changes. \\
        \hline
        \textbf{Detector Blinding Attack} & Eve blinds Bob’s detectors with bright light and controls their outputs. & Detectors forced into classical mode, responding only to Eve’s signals. & Lower detection efficiency, possible QBER increase, key fully compromised if undetected. \\
        \hline
        \textbf{Combined Attack} & Eve combines multiple strategies (e.g., RNG, detector blinding, wavelength injection). & Simultaneous exploitation of multiple vulnerabilities. & Mixed effects: increased QBER and entropy, reduced efficiency, and masked attack traces. \\
        \hline
    \end{tabular}
\end{table*}

\begin{table*}[!t]
    \centering
    \caption{Summary of Quantum Key Distribution (QKD) Metrics and Corresponding Functions}
    \label{tab:qkd_metrics}
    \begin{tabular}{|p{3cm}|p{5cm}|p{5cm}|}
        \hline
        \textbf{Metric} & \textbf{Description} & \textbf{Formula} \\
        \hline
        
        \textbf{Key Length} & The length of the final generated secret key. & N/A  \\ \hline

        \textbf{QBER} & 
        The ratio of mismatched bits after the sifting process to the total sifted key bits. Indicates the level of errors in the QKD process. &  
        \[
        QBER = \frac{\text{total mismatches}}{\text{total sifted bits}}
        \] \\ \hline

        \textbf{Measurement Entropy} & 
        Shannon entropy for matches and mismatches, measuring uncertainty in the outcomes \cite{krawec2023entropic}. &  
        \[
        H = -\sum p_i \log_2 p_i
        \] \\ \hline

        \textbf{Signal Detection Rate} & 
        Ratio of detected signal pulses to total sent signal pulses. &  
        \[
        \frac{\text{signal detections}}{\text{sent signal pulses}}
        \] \\ \hline

        \textbf{Decoy Detection Rate} & 
        Ratio of detected decoy pulses to total sent decoy pulses. &  
        \[
        \frac{\text{decoy detections}}{\text{sent decoy pulses}}
        \] \\ \hline

        \textbf{Signal Loss Rate} & 
        Ratio of lost signal pulses to total sent signal pulses. &  
        \[
        \frac{\text{lost signal pulses}}{\text{sent signal pulses}}
        \] \\ \hline

        \textbf{Decoy Loss Rate} & 
        Ratio of lost decoy pulses to total sent decoy pulses. &  
        \[
        \frac{\text{lost decoy pulses}}{\text{sent decoy pulses}}
        \] \\ \hline

        \textbf{Average Photon Transmission Time} & 
        Average time taken for a photon to be transmitted and detected. &  
        \[
        \text{Average Time} = \frac{1}{N} \sum_{i=1}^{N} T_i
        \] \\ \hline

        \textbf{Whole Key Transmission Time} & 
        Total time taken for all photons in the key exchange process. &  
        \[
        \sum T_i
        \] \\ \hline

        \textbf{Arrival Time Variance} & 
        Variance in the time taken for photon transmission, measuring fluctuations. &  
        \[
        \sigma^2 = \frac{1}{N} \sum_{i=1}^{N} (T_i - \bar{T})^2
        \] \\ \hline

        \textbf{Arrival Time Deviation} & 
        Mean absolute deviation from the average photon transmission time. &  
        \[
        \frac{1}{N} \sum_{i=1}^{N} |T_i - \bar{T}|
        \] \\
        \hline
    \end{tabular}
\end{table*}

\subsection{Attack Scenarios}
\paragraph{Normal QKD Operation}
\label{sec:normal-qkd}
In a normal  DV-QKD scenario, where there is no presence of Eve, the process of exchanging the key follows the standard BB84 protocol, where Alice randomly prepares qubits in rectilinear (horizontal/vertical) or diagonal ($\pm45^\circ$) bases, and Bob randomly chooses a basis for measurement. After transmission, they publicly compare their chosen bases and keep only the bits that match (sifting). Under these conditions, the QBER and Measurement Entropy are as close to zero as possible, indicating minimal errors and maximum predictability in the measurement outcomes. Moreover, the Signal and Decoy Detection Rates should be as close as possible to the predefined distribution ratio of signal and decoy pulses, indicating that no external interference occurs. In addition, the time-based metrics should remain stable and consistent. Any deviation from these expected values could indicate noise, losses, or an external attack on the quantum channel. Algorithm~\ref{alg:qkd_process} illustrates the pseudo-code for the normal QKD simulation. 

\begin{algorithm}[]
\caption{Normal QKD Simulation}
\label{alg:qkd_process}
\begin{algorithmic}[1]
    \State \textbf{Initialize:} Sifted key $K = []$; counters $N_{\text{sig}}, N_{\text{dec}}, N_{\text{loss,sig}}, N_{\text{loss,dec}}, N_{\text{err}}, N_{\text{det,sig}}, N_{\text{det,dec}} = 0$; time lists $T_{\text{tx}}, T_{\text{rx}} = []$
    \For{$i = 1$ to $N_{\text{trans}}$} \Comment{$N_{\text{trans}} = 50$}
        \State Randomly assign pulse type: signal ($p_{\text{sig}} = 0.7$) or decoy ($p_{\text{dec}} = 0.3$)
        \State Simulate noise and photon loss
        \If{photon is lost}
            \State Increment $N_{\text{loss,sig}}$ or $N_{\text{loss,dec}}$
            \State \textbf{continue}
        \EndIf
        \State Alice selects random bit $a \in \{0,1\}$ and basis $b_A \in \{0,1\}$
        \State Encode qubit as $|\psi\rangle = \text{Encode}(a, b_A)$
        \State Bob randomly selects measurement basis $b_B \in \{0,1\}$
        \State Bob measures: $b = \text{Measure}(|\psi\rangle, b_B)$
        \State Apply bit flip error: $b \leftarrow b \oplus \text{Bern}(p_{\text{err}})$
        \State Apply depolarization and photon shot noise to $b$
        \State $b \gets
        \begin{cases}
        b, & \text{with probability } 1 - p_{\text{depol}} \\
        \text{random}(0,1), & \text{with probability } p_{\text{depol}}
        \end{cases}$

        \State Apply photon shot noise using the Poisson model: 
        $P(k|\bar{n}) = \dfrac{\bar{n}^k e^{-\bar{n}}}{k!}$, where $k$ is the number of detected photons and $\bar{n}$ is the mean photon number
        \State Record $T_{\text{tx}}[i], T_{\text{rx}}[i]$
        \If{$b_A = b_B$}
            \State Append $(a, b)$ to $K$ (sifted key)
            \If{$a = b$}
                \State Increment $N_{\text{det,sig}}$ or $N_{\text{det,dec}}$
            \Else
                \State Increment $N_{\text{err}}$
            \EndIf
        \EndIf
    \EndFor

    \State \textbf{Calculate QBER:} \hspace{1em} $QBER = \dfrac{N_{\text{err}}}{|K|}$
    \State \textbf{Detection Rates:} \hspace{1em} $R_{\text{sig}} = \dfrac{N_{\text{det,sig}}}{N_{\text{sig}}}, \quad R_{\text{dec}} = \dfrac{N_{\text{det,dec}}}{N_{\text{dec}}}$
    \State \textbf{Loss Rates:} \hspace{1em} $L_{\text{sig}} = \dfrac{N_{\text{loss,sig}}}{N_{\text{sig}}}, \quad L_{\text{dec}} = \dfrac{N_{\text{loss,dec}}}{N_{\text{dec}}}$
    \State \textbf{Time Metrics:} \hspace{1em} $\Delta T = \dfrac{1}{|K|} \sum_{i=1}^{|K|} (T_{\text{rx}}[i] - T_{\text{tx}}[i])$
    \State \textbf{Entropy:} \hspace{1em} $H = -\sum_{x \in \{0,1\}} p_x \log_2 p_x$, where $p_x$ is the frequency of bit $x$ in the sifted key

    \State \textbf{Return:} Sifted key $K$, QBER, entropy $H$, $R_{\text{sig}}$, $R_{\text{dec}}$, $L_{\text{sig}}$, $L_{\text{dec}}$, time metrics $\Delta T$
\end{algorithmic}
\end{algorithm}

\paragraph{Intercept-and-Resend Attack} 
\label{sec:mitm-attack}
In the Intercept-and-Resend attack scenario, Eve intercepts the QKD signal, measures the transmitted photons, and then retransmits them to Bob, effectively impersonating Alice. Fundamentally, this attack exploits the direct access to the quantum channel. Eve performs a projective measurement on the in-transit qubit, effectively collapsing its quantum state. Based on her measurement outcome, she prepares and sends a new qubit to Bob. Due to the No-Cloning theorem, if Eve measures in a basis conjugate to Alice's preparation basis, she inevitably introduces state disturbances that manifest as errors in the sifted key. This process typically introduces detectable errors, but if Eve attacks only a small fraction of signals, the errors might remain below the detection threshold. Under these conditions, the QBER and Measurement Entropy significantly increase, indicating potential errors and greater unpredictability in the measurement outcomes. Furthermore, the signal and decoy detection rates should be reduced because Eve has already measured the signal or the decoy, potentially leading to external interference or tampering. Additionally, the time-based metrics are expected to increase because of the delay introduced by Eve's interception. Based on these assumptions, the simulated attack is constructed as described in Algorithm~\ref{alg:mitm_process}.
This follows the standard BB84 analysis: full-rate Intercept--Resend would yield $\approx 25\%$ QBER, so Eve targets only a fraction of pulses to avoid alarms. We therefore expect elevated QBER/entropy and mild timing overhead, matching the metrics we log. No additional simplifications apply here.

\begin{algorithm}[]
    \caption{Intercept-and-Resend Attack Simulation}
    \label{alg:mitm_process}
    \begin{algorithmic}[1]
        \State \textbf{Follows Normal QKD States 1-10}.

\State \textbf{Eve intercepts the photon:}
    \State \hspace{1em} Eve randomly selects a basis $b_E \in \{0,1\}$
    \State \hspace{1em} Eve measures the photon: $e = \text{Measure}(|\psi\rangle, b_E)$
    \State \hspace{1em} Apply bit flip error to Eve's measurement: $e \leftarrow e \oplus \text{Bern}(p_{\text{err,Eve}})$
    \State \hspace{1em} \textbf{Apply depolarization noise:}
        \[
            e \gets
            \begin{cases}
                e, & \text{with probability } 1 - p_{\text{depol,Eve}} \\
                \text{random}(0,1), & \text{with probability } p_{\text{depol,Eve}}
            \end{cases}
        \]
    \State \hspace{1em} Eve prepares a new qubit $|\psi'\rangle = \text{Encode}(e, b_E)$ and sends it to Bob
            \State \textbf{Follows Normal QKD States 11-32}

    \end{algorithmic}
\end{algorithm}

\paragraph{PNS Attack} 
\label{sec:pns-attack}

The PNS scenario follows the decoy-state model introduced by Hwang~\cite{hwang2003quantum}, in which Eve targets multi-photon pulses emitted by weak coherent sources and retains one or more photons, while allowing the remainder to reach Bob. The attack mechanism relies on the statistical nature of weak coherent laser pulses, which follow a Poisson distribution and occasionally contain more than one photon. Eve performs a Quantum Non-Demolition (QND) measurement to determine the photon number of each pulse. If she detects a multi-photon pulse, she splits off one photon to store in quantum memory and forwards the remaining photon(s) to Bob via a lossless channel. This allows her to measure the stored photon later—once the basis is announced—thereby gaining key information without inducing any bit errors. In our simulation, Eve performs photon-number–selective interception only on those multi-photon pulses; to keep the simulation computationally tractable, we do not model advanced channel-loss compensation or loss-replacement strategies that a fully resourced adversary might employ. Consequently, our traces exhibit a noticeable drop in signal/decoy detection rates, while showing little change in QBER or measurement entropy, since Eve does not measure the stored photons and therefore does not introduce additional bit errors directly. Timing metrics remain largely unaffected because the interception step introduces only a minor delay in our model. This simplified and detectable PNS variant preserves the causal link between Eve’s actions and the recorded QKD statistics (detection/loss rates, QBER, entropy), and we explicitly identify stronger covert PNS attacks that include loss compensation as out of scope for the current dataset and as limitations that will be addressed in future work.

\begin{algorithm}[]
    \caption{PNS Attack Simulation}
    \label{alg:pns_process}

    \begin{algorithmic}[1]
        \State \textbf{Follows Normal QKD States 1-9}.
        \State \textbf{Simulate photon intensity:}
            \begin{itemize}
                \item Generate a random number $r \in [0,1]$
                \item \textbf{If} $r < 0.8$, set pulse type to \textbf{single-photon}
                \item \textbf{Else}, set pulse type to \textbf{multi-photon}
            \end{itemize}
        \State \textbf{Eve's Photon Number Splitting (PNS) Attack:}
        \State Eve checks if the pulse is multi-photon.
        \If {multi-photon pulse}
            \State Eve intercepts and stores one photon.
            \State Allows remaining photons to pass without measurement.
        \Else
            \State Eve does nothing.
        \EndIf
        \State \textbf{Follows Normal QKD States 11-32}
    \end{algorithmic}
\end{algorithm}

\paragraph{Trojan Horse Attack (THA)} 
\label{sec:trojan-horse}
The THA simulated in this work follows the injection--reflection mechanism characterized in prior analyses~\cite{Lucamarini2015THA, Jain2014THA}, where Eve injects bright optical pulses into Alice’s or Bob’s QKD apparatus to extract internal information such as basis settings or key bits. This side-channel attack treats the QKD device as a passive optical target. Eve launches a bright optical probe signal into the system through the quantum channel. This probe reflects off internal components (such as phase modulators), picks up the modulation state (phase or polarization) representing the current basis/bit setting, and returns it to Eve. By analyzing this back-reflected light, Eve can read out the settings without measuring the quantum signal itself. In our implementation, Eve transmits both strong and weak probe pulses without calibrating the total optical flux at Bob’s input, forming a simplified non-calibrated-power variant. In practice, a sophisticated adversary could stabilize this flux to keep detection rates constant and avoid detection; however, that level of hardware calibration is outside the simulation scope and is explicitly treated as a limitation. This non-calibrated approach intentionally introduces measurable deviations in detection and entropy metrics—representing the operational signature of an imperfectly tuned THA while preserving the correct causal relationship between Eve’s injection and the observable QKD statistics. Specifically, injected photons occasionally reach Bob’s detectors, slightly increasing the detection rate compared to normal QKD, but leaving QBER and measurement entropy largely unchanged, as Eve does not perform measurements on the transmitted quantum states. The timing metrics also remain stable, since the injection process introduces negligible latency. These simulated outcomes are consistent with the experimental observations reported in~\cite{jain2016attacks,sushchev2024trojan}, which demonstrated that uncalibrated THA implementations create minor but detectable statistical deviations. The omission of power calibration thus reflects a deliberate modeling simplification aimed at maintaining computational tractability while preserving the physical essence of the attack. This modeling assumption remains consistent with the bounded adversary described in Section~\ref{sec:threat_model} and is acknowledged as a limitation of the present study. Based on these considerations, the THA scenario is implemented as described in Algorithm~\ref{alg:trojan_horse_process}.

\textbf{Semi-Quantitative Framework for Trojan Horse Attacks.}
Although empirical measurements for THA parameters are not yet included in this study, 
we formed a quantitative framework for modeling bias and information leakage arising from injected optical probes. 

Let $p_{\text{inj}}$ denote the probability that an adversary successfully injects a probing pulse within a communication slot, 
$\rho_{\text{ret}}$ is the effective optical return reflectivity of the internal optics, and $F_{\text{Eve}}$ is the fidelity with which the adversary infers the modulator setting from the back-reflected light. 
The guessing probability can characterize the resulting bias in key generation
\[
p_{\text{guess}} = \tfrac12(1-p_{\text{inj}}) 
+ \big(\tfrac12 + \rho_{\text{ret}}(F_{\text{Eve}}-\tfrac12)\big)p_{\text{inj}},
\]
which defines the conditional min-entropy per bit as 
\[
H_{\min}(K|E) = -\log_2 p_{\text{guess}},
\]
and the corresponding entropy loss $\delta H_{\min} = 1 - H_{\min}(K|E)$.
The expected information leakage for an $n$-bit raw key can then be upper-bounded by
\[
L_{\max} = n(1 - H_{\min}).
\]
These expressions provide an analytical framework to evaluate how probe-injection frequency, component reflectivity, 
and adversarial inference fidelity jointly influence the effective security of QKD systems under THA scenarios. 
Future simulation or experimental data can be used directly to estimate entropy reduction and leakage levels.

\makeatletter
\renewcommand{\ALG@beginalgorithmic}{\small}
\makeatother
\begin{algorithm}[]
    \caption{Trojan-Horse Attack Simulation}
    \label{alg:trojan_horse_process}

    \begin{algorithmic}[1]
        \State \textbf{Follows Normal QKD States 1-10}.
        
        \State \textbf{Eve’s Trojan-Horse Attack:}
        \State Generate a random number $r_1 \in [0,1]$
        \State \textbf{If} $r_1 < p_{\text{inject}}$ \textbf{(injection occurs):}

            \State \hspace{1em} Generate another random number $r_2 \in [0,1]$
            \State \hspace{1em} \textbf{If} $r_2 < 0.5$ \textbf{(strong pulse):}
                \State \hspace{2em} Set Bob's detection probability to high (e.g., $p_{\text{detect,strong}}$)
                \State \hspace{2em} Generate $r_3 \in [0,1]$; \textbf{if} $r_3 < 0.5$, set detected signal as mismatched
            \State \hspace{1em} \textbf{Else} \textbf{(weak pulse):}
                \State \hspace{2em} Set Bob's detection probability to low (e.g., $p_{\text{detect,weak}}$)
                \State \hspace{2em} Generate $r_3 \in [0,1]$; \textbf{if} $r_3 < 0.2$, set detected signal as mismatched
            \State \hspace{1em} Continue as normal QKD process for this pulse
        \State \textbf{Else:} No injection; proceed as normal QKD process

        \State \textbf{Follows Normal QKD States 11-32}

    \end{algorithmic}
\end{algorithm}

\paragraph{Wavelength Dependent Trojan-Horse Attack}
\label{sec:wavelength-trojan-horse}

The wavelength-dependent Trojan-Horse Attack simulated in this work targets the spectral sensitivity of QKD components rather than relying on reflected power at the legitimate wavelength, as in the standard THA. In this variant, Eve injects optical pulses at slightly offset wavelengths to exploit wavelength-dependent variations in detector efficiency and phase modulation \cite{gisin2006trojan, jain2016attacks}. This attack method takes advantage of the chromatic dispersion and wavelength-dependent coupling ratios of optical components. Optical devices such as beam splitters and modulators operate nominally at a specific design wavelength; shifting the probe wavelength allows Eve to alter splitting ratios, bypass spectral filters, or induce phase shifts that differ from the operational norm. This enables her to extract information or manipulate the system by finding spectral holes in the hardware defenses. For each injected wavelength $\lambda_{\text{attack}}$, the simulation computes the wavelength offset $\Delta\lambda = |\lambda_{\text{attack}} - \lambda_{\text{legit}}|$ and updates the detector response through efficiency $\eta(\Delta\lambda)$ and phase $\phi(\Delta\lambda)$. These wavelength-induced perturbations can reduce detection efficiency, introduce phase errors, and slightly increase photon arrival-time dispersion—effects that manifest directly in QBER, detection/loss rates, and timing metrics. While both THA variants share the same injection–reflection principle, their physical mechanisms and statistical footprints differ: the standard THA manipulates total optical flux, whereas the wavelength-dependent variant alters spectral response, producing distinct phase and timing signatures. This distinction allows the hybrid QLSTM model to recognize spectral leakage patterns without conflating them with flux-based deviations. To maintain computational feasibility, the simulation adopts a parametric model that captures the causal link between injected wavelength shifts and measurable QKD statistics, without implementing full optical modeling such as modal coupling or detector spectral calibration. More advanced, hardware-calibrated implementations of this attack are explicitly considered beyond the scope of this dataset and are listed as limitations of the present study.

\makeatletter
\renewcommand{\ALG@beginalgorithmic}{\small}
\makeatother
\begin{algorithm}[]
    \caption{Wavelength Dependent Trojan-Horse Attack Simulation}
    \label{alg:wavelength_trojan_horse_process}

    \begin{algorithmic}[1]
        \State \textbf{Follows Normal QKD States 1-10}.
        \State \textbf{Eve’s Wavelength-Dependent Trojan-Horse Attack:}
        \State Generate a random number $r_1 \in [0,1]$
        \State \textbf{If} $r_1 < p_{\text{inject}}$ \textbf{(injection occurs):}
            \State \hspace{1em} Randomly select attack wavelength $\lambda_{\text{attack}}$ from possible set
            \State \hspace{1em} Calculate wavelength difference $\Delta\lambda = |\lambda_{\text{attack}} - \lambda_{\text{legit}}|$
            \State \hspace{1em} Update detection efficiency: $\eta \leftarrow \eta_0 \cdot f(\Delta\lambda)$
            \State \hspace{1em} Update phase encoding: $\theta \leftarrow \theta + \phi(\Delta\lambda)$
            \State \hspace{1em} Generate $r_2 \in [0,1]$
            \State \hspace{1em} \textbf{If} $r_2 < p_{\text{eavesdrop}}$:
                \State \hspace{2em} Eve successfully extracts information about the key bit
            \State \hspace{1em} \textbf{If} $\Delta\lambda > 0$:
                \State \hspace{2em} With probability $p_{\text{error}}$, introduce additional measurement error at Bob's side
            \State \hspace{1em} Add wavelength-dependent delay to photon arrival time
            \State \hspace{1em} Continue as normal QKD process for this pulse
        \State \textbf{Else:} No injection; proceed as normal QKD process

        \State \textbf{Follows Normal QKD States 11-32}
    \end{algorithmic}
\end{algorithm}

\paragraph{Random Number Generator Attack}
\label{sec:rng-attack}

The \emph{Random Number Generator (RNG) Attack} targets the fundamental assumption that Alice and Bob's random choices of bits and bases are truly unpredictable and uniformly distributed. If the RNGs used in the QKD protocol are biased, patterned, or otherwise predictable—due to hardware flaws or deliberate compromise—an eavesdropper (Eve) can exploit this to increase her information about the key~\cite{jofre2011true,gerhardt2011full}. The attack capitalizes on non-random correlations or biases in the RNG output stream. Eve collects a history of public basis announcements and, if available, compromised key bits, to train a predictive model (e.g., using statistical analysis or machine learning). By forecasting the basis choices of Alice and Bob with high probability, Eve can align her intercept-resend measurements with the predicted basis, drastically reducing the error rate (QBER) that typically reveals her presence. In this attack, Eve analyzes recent output patterns from the RNG and leverages statistical bias or repeated patterns to predict future choices with non-negligible probability. If Eve can correctly guess Alice's bit and basis, she can intercept and resend the photon, introducing minimal disturbance and remaining largely undetected. The effectiveness of the attack depends on the degree of bias, the predictability of the pattern, and Eve's ability to exploit them in real time. The Algorithm~\ref{alg:rng_attack_process} describes the simulation process for this attack.

\makeatletter
\renewcommand{\ALG@beginalgorithmic}{\small}
\makeatother
\begin{algorithm}[]
    \caption{RNG Attack Simulation}
    \label{alg:rng_attack_process}
    \begin{algorithmic}[1]
        \State \textbf{Follows Normal QKD States 1-10}.
        \State Set bit bias $b_{\text{bit}}$, basis bias $b_{\text{basis}}$, pattern exploitation probability $p_{\text{pattern}}$, prediction window size $w$
        \For{each photon transmission}
            \State Alice generates a bit and basis using a compromised RNG with bias and pattern memory
            \State Record previous bits and bases for pattern prediction
            \State \textbf{If} Eve predicts bit and basis (using recent history and $p_{\text{pattern}}$):
                \State \hspace{1em} With probability $p_{\text{intercept}}$, Eve intercepts and resends photon
                \State \hspace{1em} If Eve's prediction is correct and bases match, she may introduce errors
            \State Simulate transmission noise and photon loss
            \State Bob chooses a basis using a similarly compromised RNG
            \State Bob measures received photon and records the result
        \EndFor
        \State Sift key: keep only cases where both Alice and Bob have valid detections and matching bases
        \State Calculate QBER, detection/loss rates, bias metrics, and prediction success rate
        \State \textbf{Follows Normal QKD States 11-32}
    \end{algorithmic}
\end{algorithm}

\paragraph{Detector Blinding Attack}
\label{sec:detector-blinding}

In \emph{Detector Blinding Attack}, Eve manipulates the single-photon detectors on Bob's side by sending bright continuous-wave light, forcing them into a linear (classical) mode where they no longer behave as true quantum detectors~\cite{lydersen2010hacking}. The core principle is to overwhelm the Avalanche Photodiodes (APDs) with high-intensity illumination, preventing them from operating in Geiger mode, where they are sensitive to single photons. Once blinded, the detectors are essentially turned off until a pulse with energy exceeding a high discrimination threshold arrives. Eve can then send bright 'trigger' pulses precisely when she wants a detection to occur, effectively taking full control of Bob's measurement outcomes without his knowledge. While blinded, Bob’s detectors lose their single-photon sensitivity and only respond to Eve’s specifically tailored trigger pulses, allowing Eve to control which detector clicks, and thus gain full knowledge of the key without introducing detectable errors. The attack is typically performed in bursts (blinding duration), and Eve can inject her own signals with high efficiency while the detectors are blinded. During these periods, the detection efficiency drops, and the quantum bit error rate (QBER) may increase if the attack is not perfectly synchronized. However, if performed carefully, the attack can leave the QBER and other statistics nearly unchanged, making detection extremely challenging~\cite{lydersen2010hacking, makarov2016quantum}. The Algorithm~\ref{alg:detector_blinding_process} summarizes the simulation process for this attack. We follow \cite{lydersen2010hacking}: bright CW light forces a linear regime ($\eta_{\text{blinded}} \ll \eta_{\text{normal}}$) and Eve injects triggers. In ideal synchronization, QBER may remain flat;however, occasional desynchronization yields slight QBER increases and reduced efficiency, both of which are observable in our features.

\makeatletter
\renewcommand{\ALG@beginalgorithmic}{\small}
\makeatother
\begin{algorithm}[]
    \caption{Detector Blinding Attack Simulation}
    \label{alg:detector_blinding_process}
    \begin{algorithmic}[1]
        \State \textbf{Follows Normal QKD States 1-10}.
        \State Set blinding probability $p_{\text{blind}}$, blinding duration $d_{\text{blind}}$, injection efficiency $p_{\text{inject}}$
        \State Initialize detector state as \texttt{normal} with efficiency $\eta_{\text{normal}}$
        \For{each photon transmission}
            \State Randomly select photon as signal (70\%) or decoy (30\%)
            \State \textbf{If} Eve blinds detectors ($r_1 < p_{\text{blind}}$ and not already blinded):
                \State \hspace{1em} Set detector state to \texttt{blinded} for $d_{\text{blind}}$ rounds
                \State \hspace{1em} Set efficiency to $\eta_{\text{blinded}} \ll \eta_{\text{normal}}$
            \State Alice prepares qubit (bit, basis), applies encoding
            \State Simulate transmission noise and losses
            \State \textbf{If} photon lost: record as loss, continue
            \State \textbf{If} detectors are blinded:
                \State \hspace{1em} With probability $p_{\text{inject}}$, Eve injects her own signal and chooses basis/bit
                \State \hspace{1em} Use Eve's parameters for measurement
            \State Bob measures qubit in random basis
            \State Simulate detection based on current detector efficiency
            \State \textbf{If} detection occurs:
                \State \hspace{1em} Bob records measured bit
            \State \textbf{Else:} record as loss
            \State Update blinding countdown; if expired, restore detector to normal
        \EndFor
        \State Sift key: keep only matching bases and successful detections
        \State Calculate QBER, detection/loss rates, and timing metrics
        \State {\textbf{Follows Normal QKD States 11-32}}

    \end{algorithmic}
\end{algorithm}

\paragraph{Combined Attack}
\label{sec:combined-attack}

The \emph{Combined Attack} represents a realistic and formidable threat scenario in which an eavesdropper (Eve) simultaneously employs multiple quantum-hacking strategies to maximize her chances of compromising the QKD system. In this attack, Eve leverages a combination of wavelength attacks, detector blinding, and manipulation of a random number generator (RNG). This compound strategy operates by layering distinct attack vectors to cover each other's weaknesses. For instance, the predictive power of an RNG attack can reduce the error rate inherent in intercept-resend maneuvers, while detector blinding enables Eve to control the exact timing and success of her injected pulses, masking the statistical anomalies (like loss rates) that usually accompany other active attacks. By exploiting device imperfections and protocol vulnerabilities in concert, Eve can bypass optical filters with off-wavelength photons~\cite{zhao2008quantum}, force detectors into a controllable linear mode~\cite{lydersen2010hacking}, and predict or bias random choices made by Alice and Bob~\cite{gerhardt2011full}. The synergy between these attacks allows Eve to increase her information gain while minimizing detection, as the errors and losses introduced by one attack may mask or compensate for those caused by another. The Algorithm~\ref{alg:combined_attack_process} summarizes the simulation process for this multifaceted attack.

\makeatletter
\renewcommand{\ALG@beginalgorithmic}{\small}
\makeatother
\begin{algorithm}[]
    \caption{Combined Attack Simulation}
    \label{alg:combined_attack_process}
    \begin{algorithmic}[1]
        \State \textbf{Follows Normal QKD States 1-10}.
        \State Set probabilities for wavelength attack, detector blinding, and RNG attack
        \For{each photon transmission}
            \State Randomly activate any subset of the three attacks per photon
            \State Simulate photon loss and noise
            \State Alice prepares bit and basis (biased/predictable if RNG attack active)
            \State Eve may intercept if she predicts Alice's choices (RNG attack)
            \State If wavelength attack is active, modify photon efficiency and phase
            \State If detector blinding is active, manipulate Bob's detection probability and error rate
            \State Bob chooses a basis (biased if RNG attack is active) and measures a photon
            \State Apply measurement errors from all active attacks
        \EndFor
        \State Sift key: keep only cases with valid detections and matching bases
        \State Calculate QBER, detection/loss rates, attack success rates, and timing metrics
        \State \textbf{Follows Normal QKD States 11-32}
    \end{algorithmic}
\end{algorithm}

\subsection{Generation}
\label{sec:dataset-generation}
Our dataset is constructed by executing multiple iterations of the QKD simulations discussed in the previous subsections. Each iteration simulates all eight possible scenarios (normal QKD process, Intercept-and-Resend attack, PNS attack, Trojan horse attack, wavelength-dependent Trojan horse, RNG attack, detection blinding attack, and combined attack) and captures a set of QKD metrics. These include quantum bit error rate (QBER), measurement entropy, signal and decoy detection rates, signal and decoy loss rates, photon transmission time metrics, and time-based variation metrics. These metrics provide insights into the security of the quantum channel and help detect anomalies in the process. A total of 1292 iterations were performed per scenario, resulting in a balanced dataset of 10,329 samples. Table~\ref{tab:qkd_sample} shows the first lines of the proposed dataset. Data preprocessing is a crucial step in developing any ML or DL model, as it significantly affects the model's accuracy and performance. In this context, the preprocessing framework for the proposed model contains several steps to prepare the dataset for training and evaluation. These steps include feature selection, feature transformation, dataset splitting, noise injection, and input representation.

\begin{table*}[!t]
    \centering
    \caption{Sample of the Quantum Key Distribution dataset across 8 attack scenarios}
    \label{tab:qkd_sample}
    \resizebox{\textwidth}{!}{
    \begin{tabular}{|c|c|c|c|c|c|c|c|c|c|c|}
        \hline
        \textbf{Sifted\_Key\_Length} & \textbf{QBER} & \textbf{Measurement\_entropy} & \textbf{Signal\_detection\_rate} & \textbf{Decoy\_detection\_rate} & \textbf{Avg\_Photon\_time} & \textbf{Whole\_key\_time} & \textbf{Arrival\_var} & \textbf{Arrival\_dev} & \textbf{Label} \\
        \hline
        253 & 0.0553 & 0.2981 & 0.7311 & 0.7545 & 0.0684 & 35.34 & 0.0019 & 0.0359 & normal \\
        247 & 0.3927 & 0.9661 & 0.7510 & 0.7110 & 0.1879 & 97.13 & 0.0213 & 0.1059 & mitm\_attack \\
        284 & 0.0739 & 0.3798 & 0.7112 & 0.6446 & 0.0923 & 47.62 & 0.0059 & 0.0653 & pns\_attack \\
        240 & 0.2542 & 0.8169 & 0.7437 & 0.7679 & 0.0932 & 49.02 & 0.0071 & 0.0672 & trojan\_horse\_attack \\
        311 & 0.0129 & 0.0990 & 0.8663 & 0.8738 & 0.0630 & 44.12 & 0.0024 & 0.0380 & wavelength\_dependent\_trojan\_attack \\
        454 & 0.1850 & 0.6882 & 0.9090 & 0.8943 & 0.0697 & 48.79 & 0.0040 & 0.0479 & rng\_attack \\
        142 & 0.0845 & 0.4178 & 0.4340 & 0.4265 & 0.0581 & 40.69 & 0.0096 & 0.0734 & detector\_blinding\_attack \\
        344 & 0.1744 & 0.6585 & 0.8198 & 0.8815 & 0.0653 & 45.69 & 0.0025 & 0.0374 & combined\_attack \\
        \hline
    \end{tabular}
    }
\end{table*}

\paragraph{Feature Selection} 
The dataset contains 11 features, but only 9 are used to train the proposed model, excluding the Signal Detection Rate and Decoy Detection Rate. This decision is based on the fact that the Signal Loss Rate and the Decoy Loss Rate are already included and convey the same information. Removing these features helps prevent overfitting and ensures the model generalizes well.

\paragraph{Feature Transformation} 
All feature values are standardized using Z-score normalization (StandardScaler) to ensure that they have a mean of zero and a standard deviation of one, using the following equation:
\begin{equation}
    X' = \frac{X - \mu}{\sigma}
\end{equation}
where $\mu$ represents the mean of the feature values, and $\sigma$ denotes the standard deviation. In addition, categorical labels are encoded as one-hot vectors to ensure that the model correctly interprets class information.

\paragraph{Dataset Splitting} 
The dataset is divided into training and test sets at 80\%-20\% to ensure sufficient training data while maintaining a reliable evaluation set.

\paragraph{Noise Injection} 
Gaussian noise is added to the training set to improve model robustness and enhance generalization. This noise follows a normal distribution:
\begin{equation}
    X_{\text{train noisy}} = X_{\text{train}} + \mathcal{N}(\mu, \sigma^2)
\end{equation}
where $\mathcal{N}(\mu, \sigma^2)$ represents Gaussian noise with mean $\mu = 0$ and standard deviation $\sigma = 0.05$. This ensures that the model is exposed to slightly varied training input, making it more resilient to real-world variations.

\subsubsection{Input Representation} 
The input batch size is set to 64 samples per batch, balancing training efficiency and model stability. The sequence length is fixed at 9 time steps, corresponding to the 9 selected features extracted from the QKD system.

\subsubsection{Physical Models and Dataset Rationality.}
The dataset was constructed according to the physical parameters specified in the IEEE 2021\cite{9268987} standards for BB84 systems with decoy-state.
Specifically, signal and decoy pulse intensities ($\mu_s$, $\mu_d$) and detection probabilities were embedded within the dataset generation 
process, eliminating the need to list them explicitly as independent variables. Each data record implicitly reflects the photon statistics, 
pulse repetition rate, and total number of transmitted pulses $N_{total}$ used to calculate the Quantum Bit Error Rate (QBER) through
\begin{equation}
    \text{QBER} = \frac{N_{error}}{N_{signal} + N_{decoy}}.
\end{equation}
By encoding the physical quantities into the signal and decoy detection rates, we ensure that the dataset preserves the essential 
statistical behavior of photon-level measurements. This formulation provides a physically meaningful representation of the quantum communication features, enabling a faithful yet simulation-based reproduction of the QKD channel statistics without direct access to optical hardware.

\section{Evaluation}
\label{sec:Evaluation}

This section discusses the evaluation criteria and presents the performance results of the proposed hybrid QLSTM model compared to classical CNN and LSTM models, as well as additional baseline architectures including an Artificial Neural Network (ANN) and a Recurrent Neural Network (RNN), followed by a comparative analysis with existing work in the literature. All models were evaluated using the dataset discussed in Section~\ref{sec:dataset-generation}, after applying the same preprocessing steps detailed in the same section, and trained using the same setup described in Section~\ref{subsec:Setup} to ensure fair evaluation.

\subsection{Setup}
\label{subsec:Setup}
 The three architectures QLSTM, LSTM, CNN, ANN, and RNN were trained under identical conditions.  Training was conducted on a GPU (Google Colab); both the model parameters and the data tensors were moved to the GPU device to use accelerated matrix operations and, in the case of QLSTM, faster quantum‐circuit simulations.  We used PyTorch’s \texttt{CrossEntropyLoss}, which applies a softmax on the logits followed by the multi‐class log‐loss.  Optimization was performed using the AdamW optimizer with initial learning rate $5\times10^{-4}$ and weight decay $1\times10^{-4}$.  A Cosine Annealing Warm restart scheduler with period $T_{0}=50$ epochs was applied to cyclically modulate the learning rate.  Mini‐batches of 64 samples were used throughout the training.  Early stopping with a patience of 5 epochs monitored the validation loss, halting training if no improvement was observed for 5 consecutive epochs, thereby avoiding overfitting.  Each model was trained for 10, 20, and 50 epochs under this same protocol. These parameters are summarized in Table \ref{tab:used_hyperparameters}. 

\begin{table}[H]
    \centering
    \renewcommand{\arraystretch}{1.2}
    \setlength{\tabcolsep}{8pt}
    \caption{Hyperparameters Used for QLSTM Training}
    \label{tab:used_hyperparameters}
    \begin{tabular}{|l|c|}
        \hline
        \textbf{Hyperparameter}           & \textbf{Value}                                   \\ \hline
        Optimizer                         & AdamW                                            \\ \hline
        Learning rate                     & 0.005                                           \\ \hline
        Weight decay                      & $1\times10^{-4}$                                 \\ \hline
        Scheduler                         & \makecell[l]{Cosine Annealing \\ Warm Restarts (\(T_0=50\))} \\ \hline
        Batch size                        & 64                                                                    \\ \hline
        Early stopping patience           & 5                                                \\ \hline
    \end{tabular}
\end{table}

\subsection{Metrics}
The model is evaluated based on multiple performance metrics, including accuracy, F1-score, recall, and precision, to assess its ability to detect attacks on a QKD system. These metrics are computed using the following formulas:
    \paragraph{Accuracy} It represents the overall correctness of the model by measuring the proportion of all predictions (both attack and normal) that are correctly classified.
    \begin{equation}
    \text{Accuracy} = \frac{\text{TP} + \text{TN}}{\text{TP} + \text{TN} + \text{FP} + \text{FN}}
    \end{equation}
    where TP (True Positives) is the number of attack labels that are correctly classified as attacks, TN (True Negatives) is the number of normal labels that are correctly classified as normal, FP (False Positives) is the number of normal labels that are classified as attacks, and FN (False Negatives) is the number of attack labels that are classified as normal.

    \paragraph{Precision} It measures how many predicted attacks are actually executed, reducing false positives.
    \begin{equation}
    \text{Precision} = \frac{\text{TP}}{\text{TP} + \text{FP}}
    \end{equation}

    \paragraph{Recall} It evaluates how well the model detects actual attacks, minimizing false negatives.
    \begin{equation}
    \text{Recall} = \frac{\text{TP}}{\text{TP} + \text{FN}}
    \end{equation}

    \paragraph{F1-score} It balances both precision and recall, which is useful when there is an imbalance between normal and attack labels.
    \begin{equation}
    \text{F1-score} = 2 \times \frac{\text{Precision} \times \text{Recall}}{\text{Precision} + \text{Recall}}
    \end{equation}

Furthermore, the model's performance is compared with those of classical CNN, LSTM, ANN, and RNN models, all trained under the same conditions, to assess the proposed model's effectiveness for multi-class classification. Although the CNN model is traditionally designed for spatial data, we adapted it for 1D convolutional layers by reshaping the input so that the 9 features form a 1D structure.

\section{Results}
\label{sec:Results}

\begin{figure*}[!t]
    \centering
    \includegraphics[width=\textwidth]{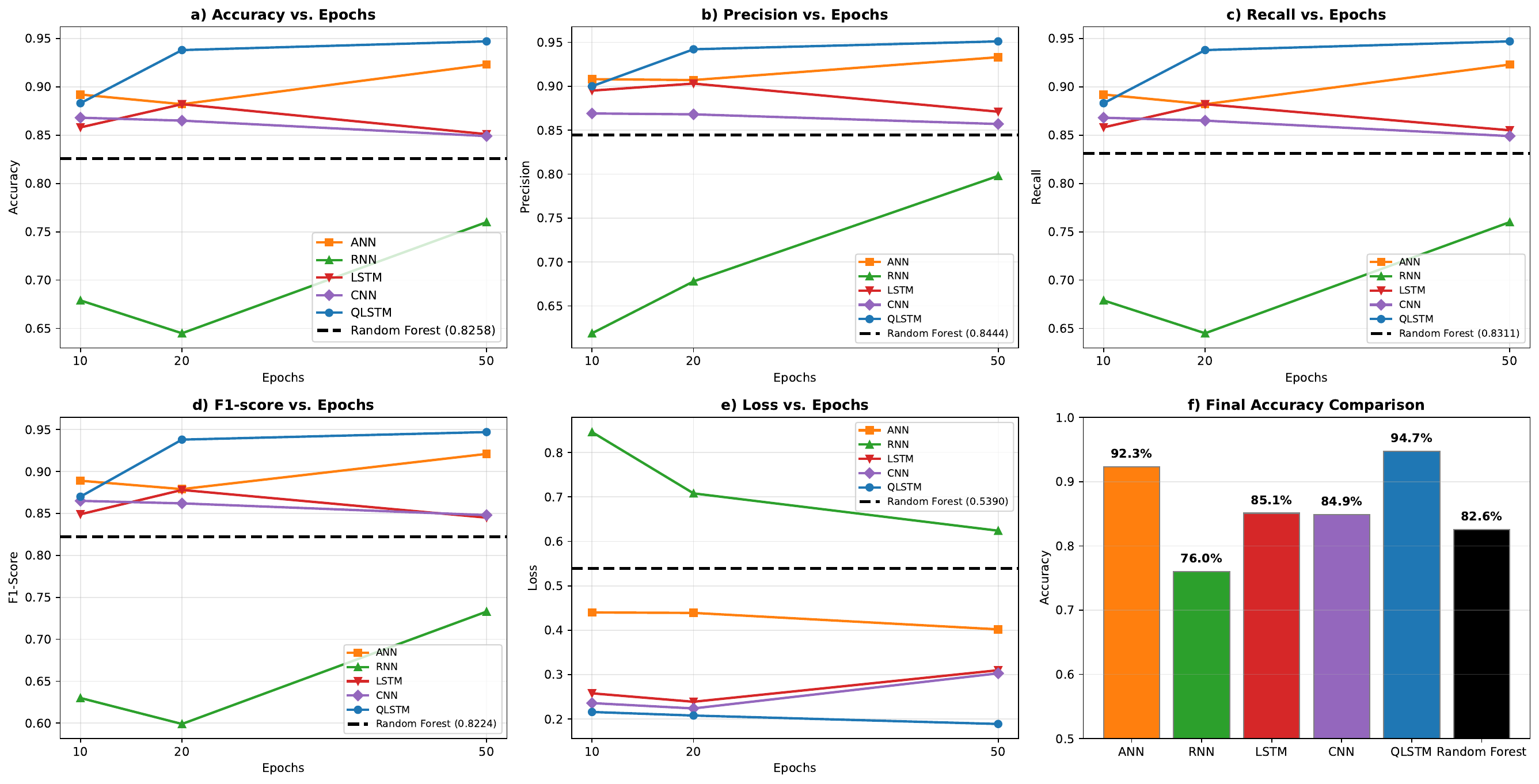}
    \caption{Performance comparison of different models: Hybrid QLSTM, LSTM, CNN, RNN, ANN, and Random Forest: a) Accuracy vs. Epochs, b) Precision vs. Epochs, c) Recall vs. Epochs, d) F1-score vs. Epochs, e) Loss vs. Epochs, and f) Final Accuracy Comparison.}
    \label{fig:model_comparisons}
\end{figure*}

The evaluation results for all models demonstrate that the proposed QLSTM model outperforms both LSTM and CNN in detecting quantum attacks on the QKD system, along with the additional ANN, Random Forest and RNN models introduced for a broader comparison. Table~\ref{tab:evaluation_metrics} shows that even in the early stages of training (10 epochs), QLSTM achieves an accuracy of 88.3\%, already higher than LSTM (85.8\%), Random Forest (82.9\%), CNN (86.8\%), and ANN (89.2\%), while significantly surpassing RNN (67.9\%). This early lead reflects QLSTM’s ability to capture temporal and statistical patterns present in quantum attack traffic, which are more challenging for classical models to extract. At 20 epochs, QLSTM continues to improve, reaching 93.8\% accuracy, and by 50 epochs it achieves 94.7\% accuracy, 95.1\% precision and 94.7\% F1-score—surpassing both classical models across all metrics. Although LSTM shows a steady but limited improvement, it peaks at 88.2\% accuracy before dropping at 50 epochs, suggesting overfitting. CNN shows a similar trend: it improves slightly until 20 epochs, then performance decreases, indicating limited ability to generalize across different attack scenarios. ANN follows a similar trend, achieving consistent but lower improvements compared to QLSTM, while RNN shows the weakest overall performance, struggling to generalize across quantum attack classes due to its limited temporal representation capacity. Figure~\ref{fig:model_comparisons} (a) illustrates how the accuracy evolves as the number of epochs increases for QLSTM, LSTM ,Random Forest,and CNN. Unlike classical models, which tend to plateau or degrade with more training, QLSTM continues to improve, reflecting its deeper learning capacity. Figures~\ref{fig:model_comparisons} (b, c, d and f) further confirm the superior performance of QLSTM, showing that it consistently achieves the highest precision, recall, and F1-score. While CNN initially maintains balanced metrics, it begins to overfit, particularly at higher epochs. Random Forest establishes a reliable baseline with high stability, yet it hits a distinguishable performance ceiling, failing to capture the complex sequential dependencies of quantum attacks as effectively as the advanced architectures. LSTM maintains a moderate balance but lacks the expressiveness to fully distinguish between quantum attack types and normal traffic in QKD processes, whereas ANN remains more robust but still underperforms QLSTM, and RNN exhibits noticeable instability across metrics.

\begin{figure*}[!t]
    \centering
    \includegraphics[width=\textwidth]{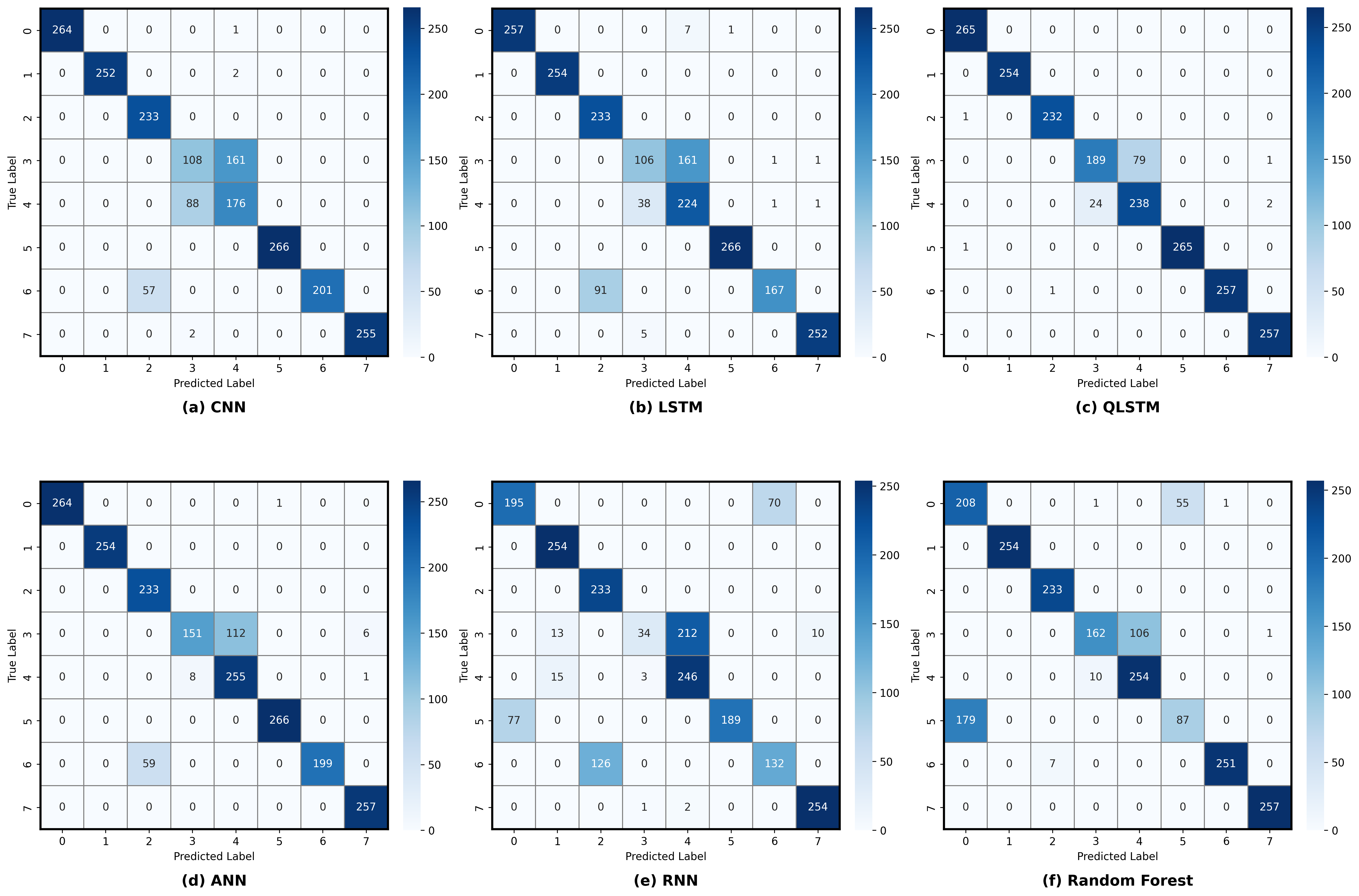}
    \caption{Confusion matrices for the evaluated models: (a) CNN, (b) LSTM, (c) Hybrid QLSTM, (d) ANN, (e) RNN, and (f) Random Forest.}
    \label{fig:conf_matrices}
\end{figure*}

To further analyze model behavior, Figure~\ref{fig:conf_matrices} presents the confusion matrices for ANN, RNN, CNN, LSTM, and QLSTM at 50 epochs. QLSTM demonstrates the most consistent and accurate classification in all eight classes. It successfully detects Combined Attack (class 0), Detector Blinding Attack (class 1), Intercept and Resend Attack (class 2), Normal (class 3), PNS Attack (class 4), RNG Attack (class 5), Trojan Horse Attack (class 6), and Wavelength Dependent Trojan Attack (class 7). Most misclassifications in QLSTM are limited to overlaps between Normal and PNS Attack, which share similar photon-level patterns. This overlap arises from their nearly identical photon-level statistics—both classes exhibit similar signal and decoy photon distributions, making the distinction subtler for any model. Adjusting the attack probabilities or photon intensity parameters during dataset generation could help reshape these class boundaries, leading to clearer separability in future experiments. In comparison, LSTM and CNN exhibit broader confusion, especially with Detector Blinding and Trojan Horse attacks. CNN also struggles more to differentiate Normal and PNS Attack, resulting in higher false positives and negatives. Although LSTM handles the Detector Blinding Attack slightly better, it still misclassifies Intercept, Resend, and Normal traffic due to their overlapping behaviors. The ANN model shows relatively stable results, performing comparably to CNN but with a slightly improved distinction between Normal and PNS Attack classes. However, it still exhibits limited confusion across high-similarity classes, reflecting its inability to fully capture quantum-feature interactions. Conversely, RNN exhibits the weakest performance among all models, with pronounced confusion across most attack classes—particularly between Combined, Intercept, and Resend, and RNG Attacks—indicating that its sequential dependencies alone are insufficient to model the quantum correlations effectively. 

These results confirm that QLSTM outperforms other methods in learning complex quantum patterns required for accurate classification in QKD environments. Table~\ref{tab:evaluation_metrics} presents the final evaluation metrics after 50 epochs of training. The QLSTM achieves better performance across all metrics considered, confirming its enhanced ability to detect and classify attacks in the QKD system.

The observed performance aligns with the theoretical expectations discussed in Section~\ref{sec:Model}.
 Hybrid QLSTM demonstrated faster convergence, higher accuracy, and stronger generalization compared to classical models, confirming the advantage of quantum-enhanced feature representation in sequential learning. These results suggest that the variational quantum circuits within the QLSTM effectively captured temporal dependencies and nonlinear correlations in the QKD parameters, leading to more stable optimization and improved discrimination between attack classes. In addition, the QLSTM’s resistance to overfitting at later epochs indicates that quantum properties such as superposition and entanglement introduce an implicit regularization effect, thereby enhancing learning efficiency. Overall, these findings validate that integrating quantum computation into recurrent architectures provides measurable benefits for complex temporal modeling in quantum key distribution security.

\begin{table}[H]
    \centering
    \caption{Performance Metrics Comparison across Different Models at Different Epochs}
    \label{tab:evaluation_metrics}
    \resizebox{\columnwidth}{!}{
    \begin{tabular}{|c|c|c|c|c|c|c|}
        \hline
        \textbf{Model} & \textbf{Epochs} & \textbf{Accuracy} & \textbf{Precision} & \textbf{Recall} & \textbf{F1-Score} & \textbf{Test Loss} \\
        \hline
        \multirow{3}{*}{ANN}  
        & 10  & 89.2\% & 90.8\% & 89.2\% & 88.9\% & 0.440\\
        & 20  & 88.2\% & 90.7\% & 88.2\% & 87.9\% & 0.439\\
        & 50  & 92.3\% & 93.3\% & 92.3\% & 92.1\% & 0.402\\
        \hline
        \multirow{3}{*}{RNN}  
        & 10  & 67.9\% & 61.9\% & 67.9\% & 63.0\% & 0.846\\
        & 20  & 64.5\% & 67.8\% & 64.5\% & 59.9\% & 0.708\\
        & 50  & 76.0\% & 79.8\% & 76.0\% & 73.3\% & 0.624\\
        \hline
        \multirow{3}{*}{LSTM}  
        & 10  & 85.8\% & 89.5\% & 85.8\% & 84.9\% & 0.258\\
        & 20  & 88.2\% & 90.3\% & 88.2\% & 87.8\% & 0.239\\
        & 50  & 85.1\% & 87.1\% & 85.5\% & 84.5\% & 0.310\\
        \hline
        \multirow{3}{*}{CNN}  
        & 10  & 86.8\% & 86.9\% & 86.8\% & 86.5\% & 0.236\\
        & 20  & 86.5\% & 86.8\% & 86.5\% & 86.2\% & 0.224\\
        & 50  & 84.9\% & 85.7\% & 84.9\% & 84.8\% & 0.303\\
        \hline
        \multirow{3}{*}{Random Forest}
        & 500 (Trees) & 82.6\% & 84.4\% & 83.1\% & 82.2\% & 0.5390\\
        & 1000 (Trees) & 81.8\% & 83.4\% & 82.4\% & 81.3\% & 0.5367\\
        & 10000 (Trees) & 81.5\% & 83.0\% & 82.0\% & 80.5\% & 0.5416\\
        \hline
        \multirow{3}{*}{QLSTM}  
        & 10  & 88.3\% & 90.0\% & 88.3\% & 87.0\% & 0.216\\
        & 20  & 93.8\% & 94.2\% & 93.8\% & 93.8\% & 0.208\\
        & 50  & \textbf{94.7\%} & \textbf{95.1\%} & \textbf{94.7\%} & \textbf{94.7\%} & \textbf{0.189}\\
        \hline
    \end{tabular}
    }
\end{table}

To complement the performance comparison across ANN, CNN, RNN, LSTM, and the proposed Hybrid QLSTM, we also benchmarked the computational requirements of each model under identical hardware and training settings. All experiments were executed using the same batch size, optimizer, learning rate, early-stopping configuration, and GPU (NVIDIA A100). Table~\ref{tab:resource_benchmark} reports the number of trainable parameters, average wall-clock time per epoch, total training time for 50 epochs, inference latency per sample, and maximum GPU memory usage for each architecture.

\begin{table}[H]
    \centering
    \caption {Computational Resource and Training-Time Benchmarking Across Models}
    \label{tab:resource_benchmark}
    \resizebox{\columnwidth}{!}{
    \begin{tabular}{|c|c|c|c|c|c|}
        \hline
        \textbf{Model} & \textbf{Params} & \textbf{Time/Epoch (s)} & \textbf{50-Epoch Time (s)} & \textbf{Infer/Sample (ms)} & \textbf{GPU Mem (MB)} \\
        \hline

        \multirow{1}{*}{ANN}
        & $1.4\times10^{4}$
        & 0.10
        & 5.0
        & 0.02
        & 120 \\
        \hline

        \multirow{1}{*}{RNN}
        & $1.1\times10^{4}$
        & 0.09
        & 4.5
        & 0.02
        & 100 \\
        \hline

        \multirow{1}{*}{CNN}
        & $2.1\times10^{4}$
        & 0.18
        & 9.0
        & 0.03
        & 150 \\
        \hline

        \multirow{1}{*}{LSTM}
        & $4.3\times10^{4}$
        & 0.30
        & 15.0
        & 0.05
        & 190 \\
        \hline
        \multirow{1}{*}{Random Forest}
        & $5.1\times10^{5}$
        & -
        & 8.9
        & 0.07
        & 0 \\
        \hline

        \multirow{1}{*}{\textbf{QLSTM}}
        & \textbf{$3.9\times10^{3} + 72$}
        & \textbf{0.41}
        & \textbf{20.5}
        & \textbf{0.06}
        & \textbf{225} \\
        \hline

    \end{tabular}
}
\end{table}

As shown in Table~\ref{tab:resource_benchmark}, the Hybrid QLSTM exhibits the highest time per epoch (0.41\,s) and the largest overall training time (20.5\,s for 50 epochs). This overhead is expected, as each forward and backward pass requires evaluating variational quantum circuits, which involve parameterized unitary transformations and repeated sampling. Despite this added cost, the Hybrid QLSTM maintains a notably lower number of classical parameters than LSTM and CNN and achieves the best accuracy, precision, recall, and F1-score among all models (Table~VIII). These results indicate that QLSTM offers a favorable trade-off: a modest increase in computational cost in exchange for significantly improved detection performance in QKD intrusion-detection scenarios.

\section{Conclusion}
This work addressed the security vulnerabilities that persist in practical QKD implementations by introducing a hybrid quantum–classical intrusion detection system based on a QLSTM architecture. Although QKD offers theoretically unbreakable security at the protocol level, its physical implementations remain vulnerable to a variety of sophisticated attacks—including Intercept–Resend, PNS, Trojan-Horse, and wavelength-dependent Trojan-Horse attacks, RNG tampering, detector blinding, and composite strategies. To mitigate these threats, our approach integrates quantum-enhanced temporal learning within a deep learning framework to strengthen the resilience of QKD networks. A key contribution of this work is the construction of a realistic and comprehensive QKD security dataset, which addresses the lack of publicly available resources for QKD intrusion detection. The dataset incorporates essential quantum communication and security parameters—QBER, entropy metrics, decoy-state statistics, photon count distributions, and timing-based indicators—to emulate real-world QKD operational behavior. Multiple attack scenarios were included to ensure rigorous and diverse model evaluation. Using this dataset, we developed a QLSTM-based IDS capable of capturing temporal, statistical, and quantum-derived signatures for accurate detection of abnormal behavior in QKD traffic. Benchmarking against strong classical baselines—including LSTM and CNN models—demonstrated that the proposed QLSTM achieves a detection accuracy of 94.7\% after 50 epochs and exhibits superior generalization capabilities. Although the Random Forest model established a robust baseline with high stability and computational efficiency (achieving $\sim$82.6\% accuracy), it reached a distinguishable performance ceiling, confirming that capturing the complex sequential dependencies of sophisticated quantum attacks requires the advanced temporal expressiveness found in quantum-enhanced architectures. These findings highlight the growing potential of hybrid quantum–classical models to improve security monitoring in emerging quantum communication infrastructures.

Despite these promising results, several limitations remain. First, the current training pipeline uses fixed hyperparameters; more systematic hyperparameter optimization could further enhance model stability and generalization. Second, although the dataset includes eight representative attack types, the attack space in practical QKD networks is broader. Extending the dataset with additional attack models would support a more comprehensive robustness analysis. Third, the dataset used in this study simulates semi-realistic QKD behavior via simplified, simulation-friendly variants of known attacks. For example, the implemented PNS attack omits optical-loss compensation steps, and the Trojan-Horse variants employ non-calibrated power settings and wavelength models rather than full hardware-level emulation. Fourth, although algorithmic simulations of the QLSTM framework demonstrate strong performance, current Noisy Intermediate-Scale Quantum (NISQ) hardware lacks the fidelity, qubit connectivity, and coherence times required for accurate runtime benchmarking. As quantum processors mature, evaluating QLSTM on real devices will help quantify the true computational overhead. Fifth, exploring alternative architectures—such as GRUs, quantum-enhanced GRUs, hybrid encoder–decoder models, or attention-based QML variants—may yield improvements in accuracy and efficiency. Sixth, comparing QLSTM performance with recent state-of-the-art QKD intrusion detection methods will provide deeper insights into its relative strengths and limitations.
Lastly, future work will incorporate quantitative evaluation of Trojan-Horse effects by computing parameters such as $(p_{\text{inj}}, \rho_{\text{ret}}, F_{\text{Eve}})$ through optical simulations or controlled laboratory experiments. This will enable direct validation of the proposed min-entropy and leakage bounds and further enhance the QLSTM's sensitivity to side-channel deviations in the physical layer. Furthermore, integrating calibrated flux control, adaptive loss modeling, and hardware-level optical parameters to narrow the gap between simulated and experimental conditions.


\ifCLASSOPTIONcaptionsoff
  \newpage
\fi

\bibliographystyle{IEEEtran}
\bibliography{Main_paper_bib}


\end{document}